\documentclass[journal=jacsat,manuscript=article]{achemso}

\usepackage{chemformula} 
\usepackage[T1]{fontenc} 
\usepackage{amssymb}
\usepackage{xcolor}

\author{Jakob D. Asmussen}
\affiliation{Department of Physics and Astronomy,
	Aarhus University, Denmark}
\author{Rupert Michiels}
\affiliation{Institute of Physics, University of Freiburg, Germany}
\author{Ulrich Bangert}
\affiliation{Institute of Physics, University of Freiburg, Germany}
\author{Nicolas Sisourat}
\affiliation{Sorbonne Universit\'e, CNRS, Laboratoire de Chimie Physique Mati\`ere et Rayonnement, France}
\author{Marcel Binz}
\affiliation{Institute of Physics, University of Freiburg, Germany}
\author{Lukas Bruder}
\affiliation{Institute of Physics, University of Freiburg, Germany}
\author{Miltcho Danailov}
\affiliation{Elettra-Sincrotrone Trieste S.C.p.A., Italy}
\author{Michele Di Fraia}
\affiliation{Elettra-Sincrotrone Trieste S.C.p.A., Italy}
\author{Raimund Feifel}
\affiliation{Department of Physics, University of Gothenburg, Sweden}
\author{Luca Giannessi}
\affiliation{Elettra-Sincrotrone Trieste S.C.p.A., Italy}
\author{Oksana Plekan}
\affiliation{Elettra-Sincrotrone Trieste S.C.p.A., Italy}
\author{Kevin C. Prince}
\affiliation{Elettra-Sincrotrone Trieste S.C.p.A., Italy}
\author{Richard J. Squibb}
\affiliation{Department of Physics, University of Gothenburg, Sweden}
\author{Daniel Uhl}
\affiliation{Institute of Physics, University of Freiburg, Germany}
\author{Andreas Wituschek}
\affiliation{Institute of Physics, University of Freiburg, Germany}
\author{Marco Zangrando}
\affiliation{Elettra-Sincrotrone Trieste S.C.p.A., Italy}
\author{Carlo Callegari}
\affiliation{Elettra-Sincrotrone Trieste S.C.p.A., Italy}
\author{Frank Stienkemeier}
\affiliation{Institute of Physics, University of Freiburg, Germany}
\author{Marcel Mudrich}
\email{mudrich@phys.au.dk}
\affiliation{Department of Physics and Astronomy,
	Aarhus University, Denmark}

\title{Time-resolved Ultrafast Interatomic Coulombic Decay in Superexcited Sodium-doped Helium Nanodroplets}


\begin{document}
	

	\begin{abstract}
		The autoionization dynamics of superexcited superfluid He nanodroplets doped with Na atoms is studied by extreme-ultraviolet (XUV) time-resolved electron spectroscopy. Following excitation into the higher-lying droplet absorption band, the droplet relaxes into the lowest metastable atomic $1s2s$ $^{1,\,3}$S states from which Interatomic Coulombic Decay (ICD) takes places either between two excited He atoms or between an excited He atom and a Na atom attached to the droplet surface. Four main ICD channels are identified and their time constants are determined by varying the delay between the XUV pulse and a UV pulse that ionizes the initial excited state and thereby quenches ICD. The time constants for the different channels all fall in the range $\sim$1~ps indicating that the ICD dynamics are mainly determined by the droplet environment. A periodic modulation of the transient ICD signals is tentatively attributed to the oscillation of the bubble forming around the localized He excitation. The ICD efficiency depends on the total number of excited states in a droplet rather than the density of excited states pointing to a collective enhancement of ICD.
        \end{abstract}
      
Nonlocal electronic decay processes play an important role in the complex ionization dynamics of weakly-bound matter irradiated by ionizing radiation. Among these processes, Interatomic (or Intermolecular) Coulombic Decay (ICD)~\cite{Cederbaum1997a} has generated considerable interest due to its general relevance including biological systems~\cite{Ouchi,hergenhahn2011interatomic,Jahnke2020}. In ICD, a valence or core shell-excited atom (or molecule) de-excites and transfers its energy to a neighboring atom which is ionized. The release of a low-energy electron in the process makes ICD efficient in causing radiation damage to DNA since slow electrons have been found to be particularly genotoxic~\cite{Alizadeh2013,Schwestka2019}. Both theory and experiment have demonstrated that ICD occurs in hydrated biosystems making the process likely to be present in biological tissue~\cite{Stoychev2011,Ren2018}. The first time-resolved experiment of ICD in prototypical neon dimers revealed an ultrashort lifetime $\sim 150$~fs~\cite{Schnorr2013}.
Studying ICD in larger molecular and condensed-phase systems becomes challenging because many other interactions of the initial and final states with the surrounding environment are possible such as scattering of the emitted electron and ion. In this regard, weakly-bound nanosystems such as van der Waals clusters are very valuable for obtaining a more complete picture of the ionization dynamics occurring in the condensed phase. Clusters and nanodroplets can be generated with a high degree of control of their structure and composition, and their finite size allows us to measure emitted electrons, ions and even photons~\cite{knie2014detecting} with high sensitivity. Superfluid helium (He) nanodroplets are ideal model systems to study electronic relaxation and ionization dynamics, not only because of the extremely weak interatomic binding of the He atoms to each other and to dopants, but also because the simple electronic structure of the He atom greatly facilitates the interpretation of electron spectra~\cite{Mudrich2014,kelbg2019auger,Ovcharenko2020a}. ICD and ICD-like decay mechanisms have been studied in both pure He droplets~\cite{Ovcharenko2014,Laforge2014a,Shcherbinin2017a,Wiegandt2019,Ovcharenko2020a} and in droplets doped with various metal atoms and molecules~\cite{Frochtenicht1996,Wang2008,Buchta2013,Laforge2016c,Shcherbinin2018b,BenLtaief2019a,Laforge2019a,BenLtaief2020,mandal2020penning}. In the case where atomic or molecular impurities are ionized through resonant valence-shell excitation of the He droplet, the ICD process resembles what traditionally has been termed Penning ionization by analogy to collisional ionization in a crossed beam experiments involving metastable rare gas atoms~\cite{Siska1993,Jahnke2020,BenLtaief2020}. 

When an excited atom is created in the droplet, a void bubble forms around the atom due to Pauli repulsion from the surrounding superfluid environment~\cite{Hansen1972,VonHaeften2002,Mudrich2020}. In a recent time-resolved experiment, we measured the dynamics of ICD in pure He nanodroplets by multiply exciting the droplets on the strongest absorption band associated mainly with the atomic $1s2p$ state~\cite{Joppien1993}. We found ICD to be very fast due to the strong attraction between the two excited atoms as the two void bubbles formed around them merge into one. In the present paper, we present time-resolved photoelectron spectra of He nanodroplets doped with sodium (Na) atoms excited into the higher-lying droplet band associated with the atomic $1s3p$ and $1s4p$ states of He. From the reappearance following depletion of the excited He states with a UV probe laser, we infer the effective lifetime of ICD for both the system of two excited He atoms in the droplet and for an excited He atom and a Na atom after relaxation from the superexcited state into the metastable $1s2s\,^1S$ and $^3S$ states. This is to our knowledge the first time-resolved study of ICD in a large heterogeneous cluster.  For a discussion of the relaxation dynamics of superexcited pure He nanodroplets, we refer to our recent paper~\cite{Asmussen2021}.
\begin{figure}[t]
	\centering
	\includegraphics[width = \columnwidth]{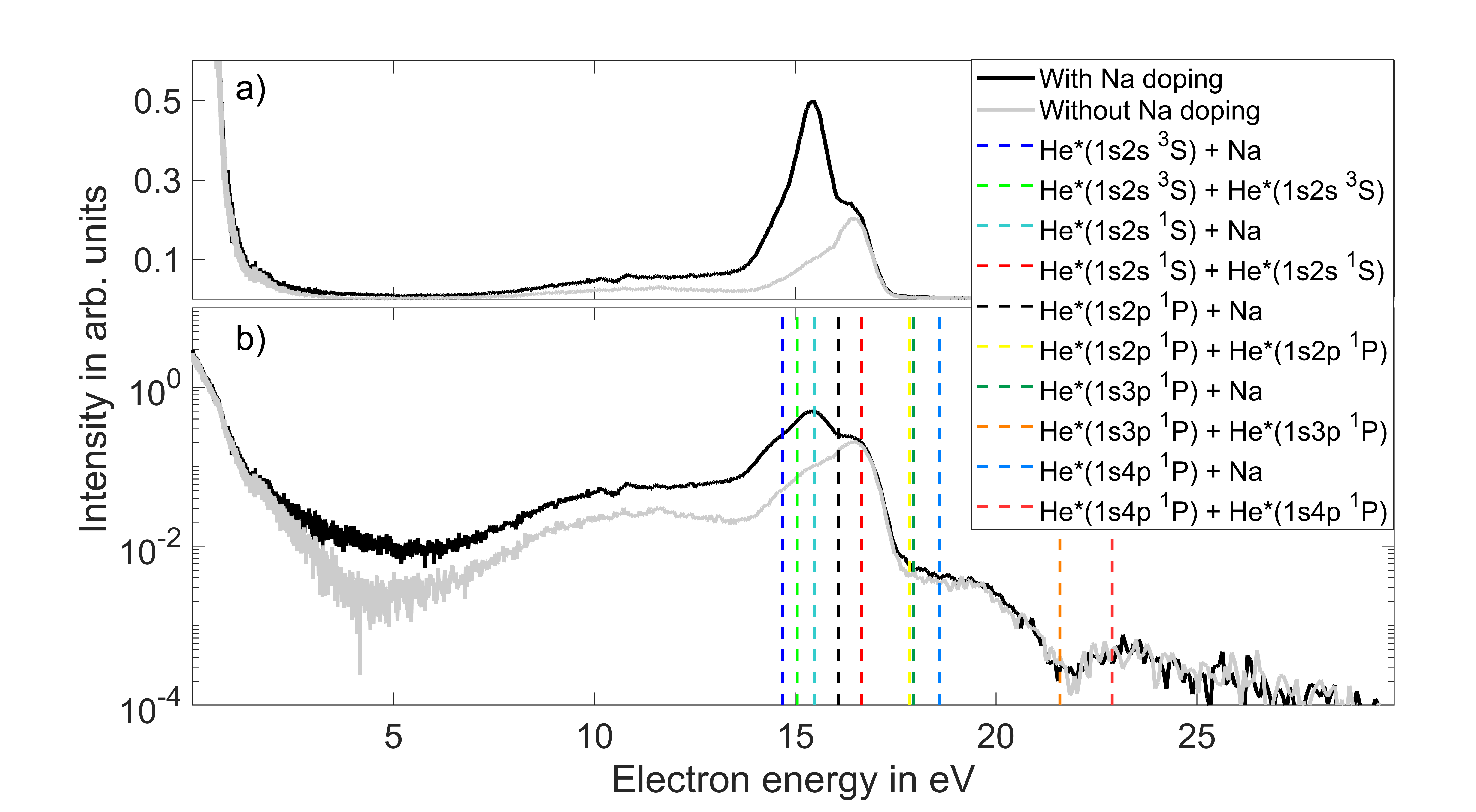}
	\caption{Static photoelectron spectra, on a linear (a) and logarithmic (b) scale, of He nanodroplets measured by multiply exciting the droplets at a photon energy of 23.7~eV. The droplets were either pure (grey line) or doped with Na atoms (black line). The dashed lines indicate the electron energies calculated for ICD from the energy levels of the unperturbed atoms\cite{NIST_He}. The two spectra were measured for droplets with an average size of $4.1\times10^4$~atoms and a FEL intensity of $1.7\times 10^{10}$~Wcm$^{-2}$.} 
	\label{1D_static}
\end{figure}
\begin{figure}[h]
	\centering
	\includegraphics[width = 0.5\columnwidth]{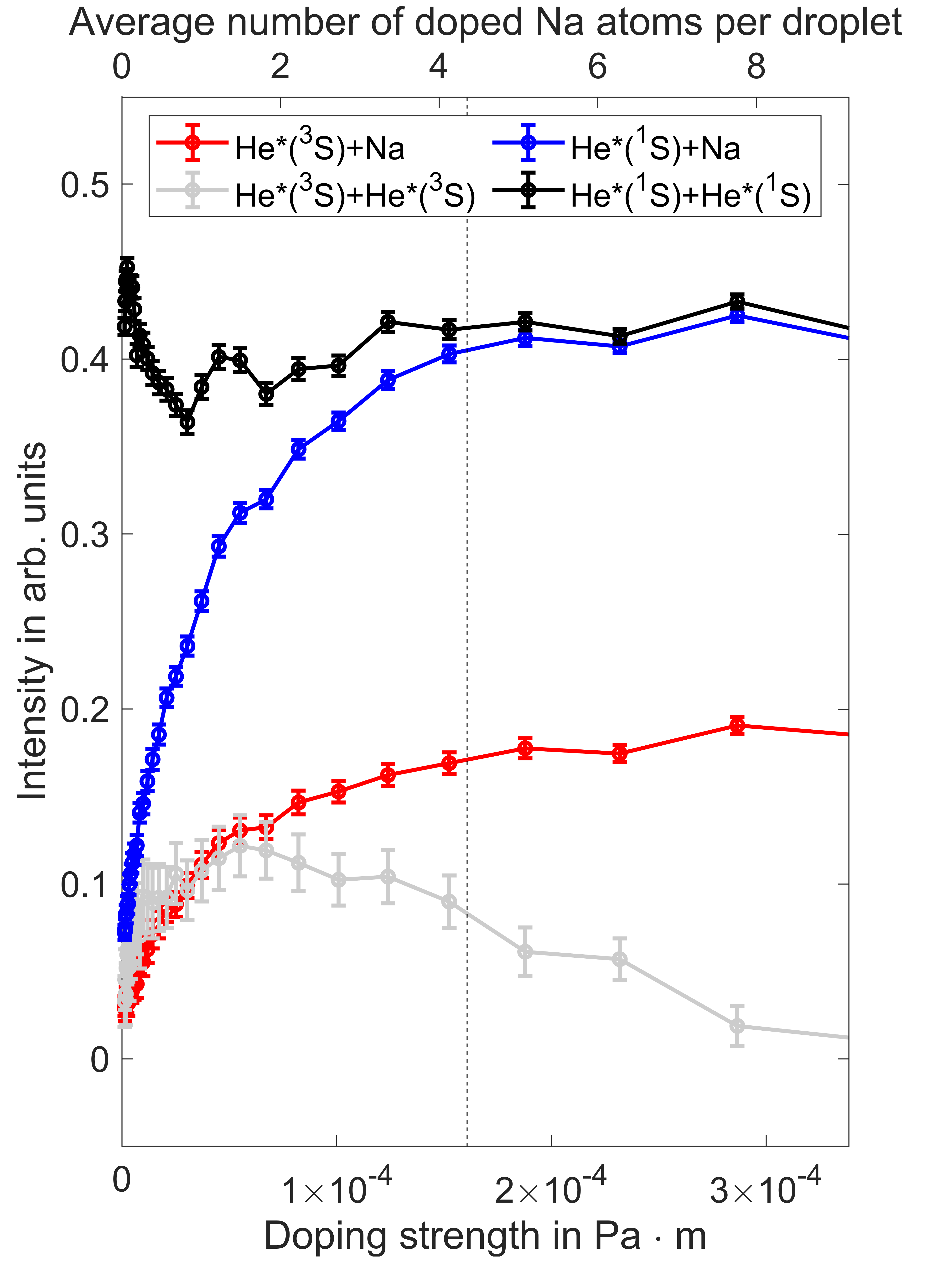}
	\caption{Yield of the 4 main ICD channels involving the lowest excited states $1s2s\,^{1,\,3}S$ as a function of doping strength (in units of tabulated Na vapor pressure times the length of the doping region). The average number of doped Na atoms (top axis) is calculated from the Poissonian distribution~\cite{Toennies2003}. The average droplet size is $4.1\times10^4$~He atoms and the FEL intensity is $1.7\times 10^{10}$~Wcm$^{-2}$. The doping strength was varied by heating the doping cell from 120$^\circ$C to 220\,$^\circ$C. The error bars show the standard deviation from the least-square regression of the multi-Gaussian fit model.   } 
	\label{Na_doping}
\end{figure}

We start by presenting the results of static measurements (XUV only) for various doping levels of Na. Figure \ref{1D_static} shows photoelectron spectra (PES) of pure and Na-doped He nanodroplets of average size $4.1\times10^4$~atoms per droplet and a FEL intensity of $1.7\times 10^{10}$~Wcm$^{-2}$. The He nanodroplets were doped with Na by heating the oven cell (see Experimental methods) to 190\,$^\circ$C corresponding to an average number of dopants per droplets around 4, assuming that the doping follows a Poissonian distribution and that the pick-up cross-section of the droplet is equal the geometrical cross-section~\cite{Toennies2004a}. As for alkali metals the doping statistics slightly deviates from Poissonian due to the desorption of dopants during the pick-up process, the estimated number of dopants should be considered as an upper bound~\cite{Bunermann2011}. The most intense feature in the spectra is given by the zero kinetic energy (ZEKE) electrons, which comes from autoionization through formation of He$_2^+$ since we excite above the adiabatic ionization potential of the droplet~\cite{Frochtenicht1996,Peterka2003,Peterka2007}.
The spectra show that the main contributions to ICD are from the metastable $1s2s\,^1S$ and $^3S$ excited states after electronic relaxation from the initial superexcited state, matching previous measurements for pure and doped He nanodroplets~\cite{Ovcharenko2020a,BenLtaief2020}. Thus, electronic relaxation~\cite{Mudrich2020,Asmussen2021} proceeds much faster than ICD. The four main channels are ICD between two equally excited He atoms ($^{1,\,3}S$) and ICD of Na through relaxation of an excited He atom ($^{1,\,3}S$) into the ground state. ICD from highly-excited states is almost two orders of magnitude less intense, and the primary channel for this is ICD from the $1s3p\,^1P$ state. The excitation into the $1s3p$ Rydberg state configuration of He most likely happens at the surface of the droplet where the He excitation is nearly unperturbed and therefore the spectral overlap with the XUV pulse is highest~\cite{Kornilov2011a}. 
The contribution to the PES at electron energies in the range 6-14~eV stems from ICD between an excimer (He$_2^*$) and another excited helium atom or Na dopant~\cite{Nijjar2018,Laforge2021}, and possibly double-ICD where the relaxation of a single He atom leads to ionization of two Na atoms~\cite{Laforge2019a}. Since these different channels all lead to broad maxima in the PES which we cannot disentangle, and the resulting broad feature showed no significant dependence on the pump-probe delay, we refrain from further analyzing this part of the spectra. 

\begin{figure}[t]
	\centering
	\includegraphics[width = \columnwidth]{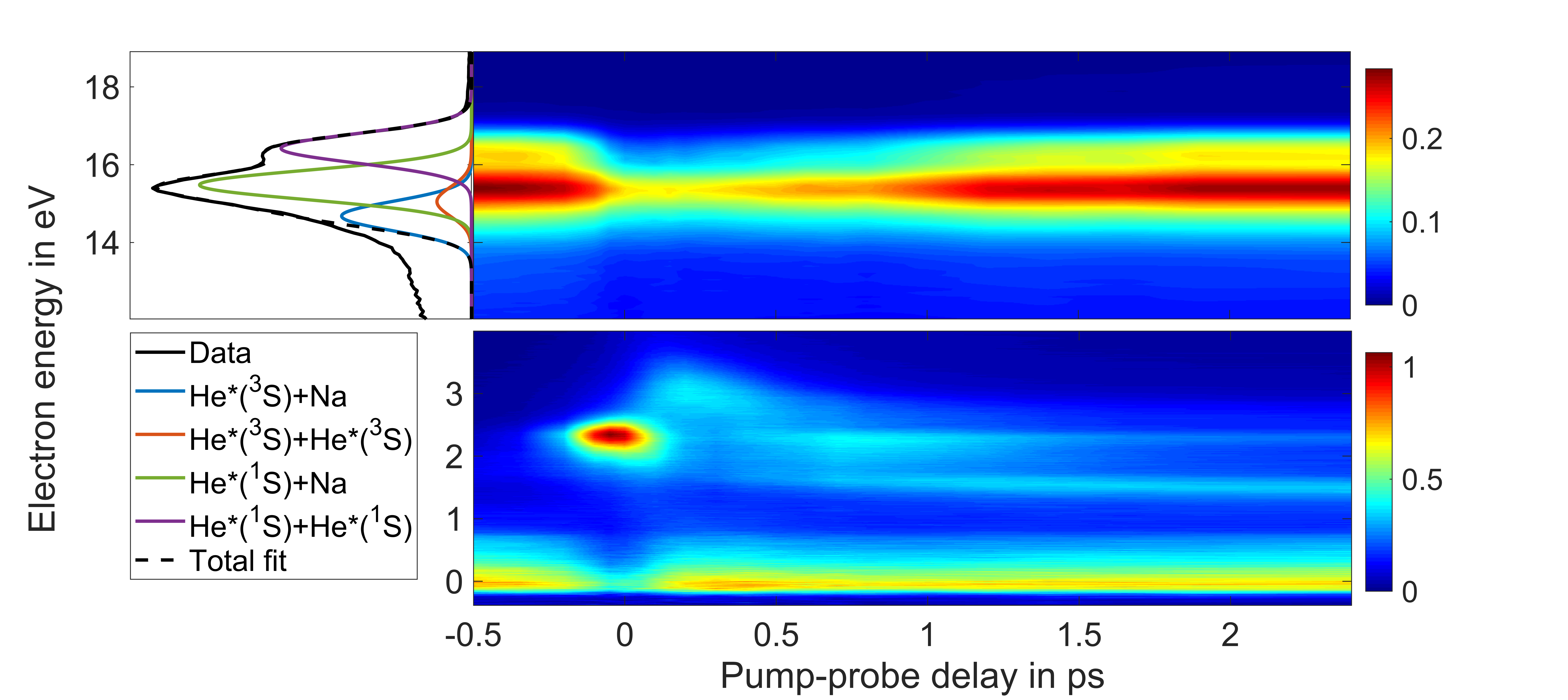}
	\caption{Time-resolved electron spectra for Na-doped He nanodroplets containing on average $1.5\times10^5$~He atoms. The lower part shows photoelectrons due to ionization of the excited He atoms with either one or two photons of the UV probe~\cite{Asmussen2021}. The upper part shows electrons generated by ICD which is quenched around 0~ps pump-probe delay where the excited state is photoionized before ICD has happened. The panel on the top left shows an electron spectrum at negative delay (UV first), where the ICD process is unperturbed. The yield of the different ICD channels is determined from a multi-Gaussian fit (colored lines). The spectra were recorded at a FEL intensity of $1.0\times 10^{10}$~Wcm$^{-2}$ and a UV probe intensity of $2.6\times10^{11}$~Wcm$^{-2}$.  }
	\label{TRPES}
\end{figure}

To further analyze the ICD dynamics in Na-doped He nanodroplets, we use a multi-Gaussian fit model consisting of the sum of four Gaussians, one for each of the four main channels. Figure \ref{Na_doping} shows the peak area of the four different Gaussian peaks as a function of doping strength. 
The peak widths for all four Gaussians were kept equal and treated as a variable parameter, and the position of each peak was set to the calculated value based on the atomic ionization energies of He and Na and the atomic level energies of excited states.\cite{NIST_He} Only the He*($^1$S)+He*($^1$S) ICD peak was shifted from the calculated value ($16.65$~eV) to $16.42$~eV to improve the convergence of the fit model. A shift towards lower kinetic energy has previously been reported for ICD electrons detected in coincidence with ions~\cite{Buchta2013,BenLtaief2019a}. 
Due to the close spacing of the peaks relative to their widths (see figure \ref{TRPES}, top-left corner), a significant uncertainty of the fitted peak intensities should be taken into account in the following analysis, especially for the least dominant peak (He*($^3$S)+He*($^3$S)). Since ICD involving the He*($^3$S) state is clearly present in the spectra, this channel is nevertheless included in the fit model. 

In Figure \ref{Na_doping}, the doping strength was varied by heating the oven cell from 120\,$^\circ$C to 220\,$^\circ$C at constant initial droplet size and FEL intensity as in Figure \ref{1D_static}. The ionization of Na by ICD rises as expected with increasing doping strength while the He*+He* ICD contribution remains nearly constant for all doping strengths indicating that depletion and deflection of the He droplet beam by collisions with Na atoms in the vapor are negligible at these doping pressures. Na, being an alkali metal, is heliophobic and will form a cluster at the surface of the droplet. Only at higher doping strengths where the size of the Na cluster exceeds 21 atoms, will the dopant cluster sink into the droplet~\cite{AnDerLan2011}. The intensity of ICD leading to ionization of Na saturates at a doping strength around $1.6\times10^{-4}$~Pa$\cdot$m, which corresponds to the doping strength at which the contribution of undoped He droplets tends to zero assuming a Poissonian distribution. Note that for doping with alkali metal, extensive simulations have shown that the contribution of undoped droplets never becomes zero, but converges to $\sim$10\% of its initial value~\cite{Bunermann2011}. The saturation of the He*+Na ICD signal for high doping levels supports the general concept that alkali-metal atoms picked up by He droplets aggregate to form a cluster which effectively undergoes ICD as a single particle. The time-resolved measurements presented in the following were carried out at this doping level (190\,$^\circ$C).
\begin{figure}[h]
	\centering
	\includegraphics[width = 0.5\columnwidth]{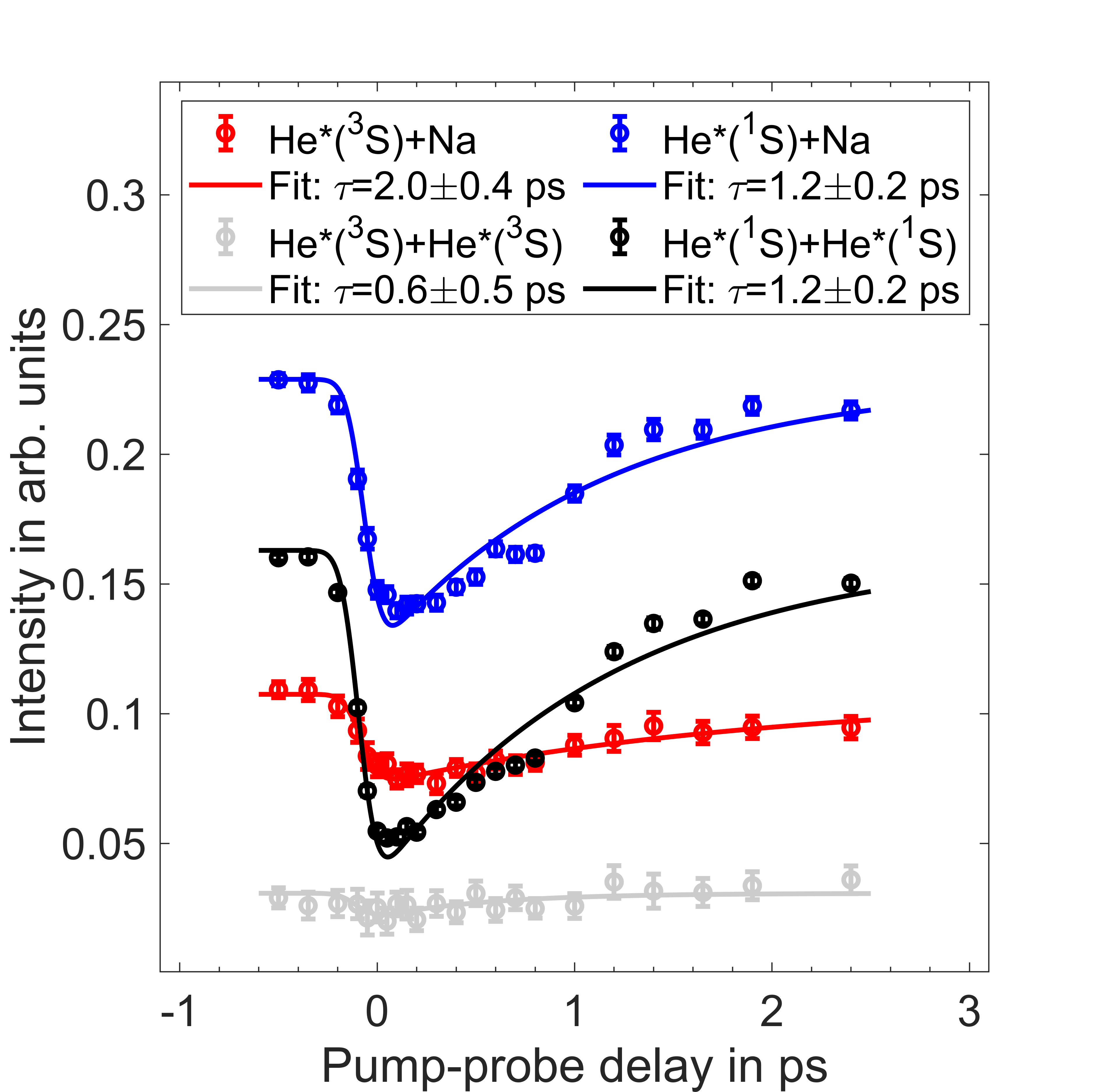}
	\caption{Electron yield from the 4 ICD channels measured at different pump-probe delays for Na-doped droplets containing on average $1.5\times10^5$~He atoms and measured with a FEL intensity of $1.0\times 10^{10}$~Wcm$^{-2}$ and an UV probe intensity of $2.6\times10^{11}$~Wcm$^{-2}$. The depletion and exponential rise of each channel is fitted by an exponential-decay model (see the main text for details) to determine the ICD time constant. The error bars on the data points show the standard deviation from the least-square regression of the multi-Gaussian fit model, and the resulted ICD lifetimes from the exponential decay model are given in the legend with standard deviation. }
	\label{depletion}
\end{figure}
\begin{figure}[h]
	\centering
	\includegraphics[width = 0.5\columnwidth]{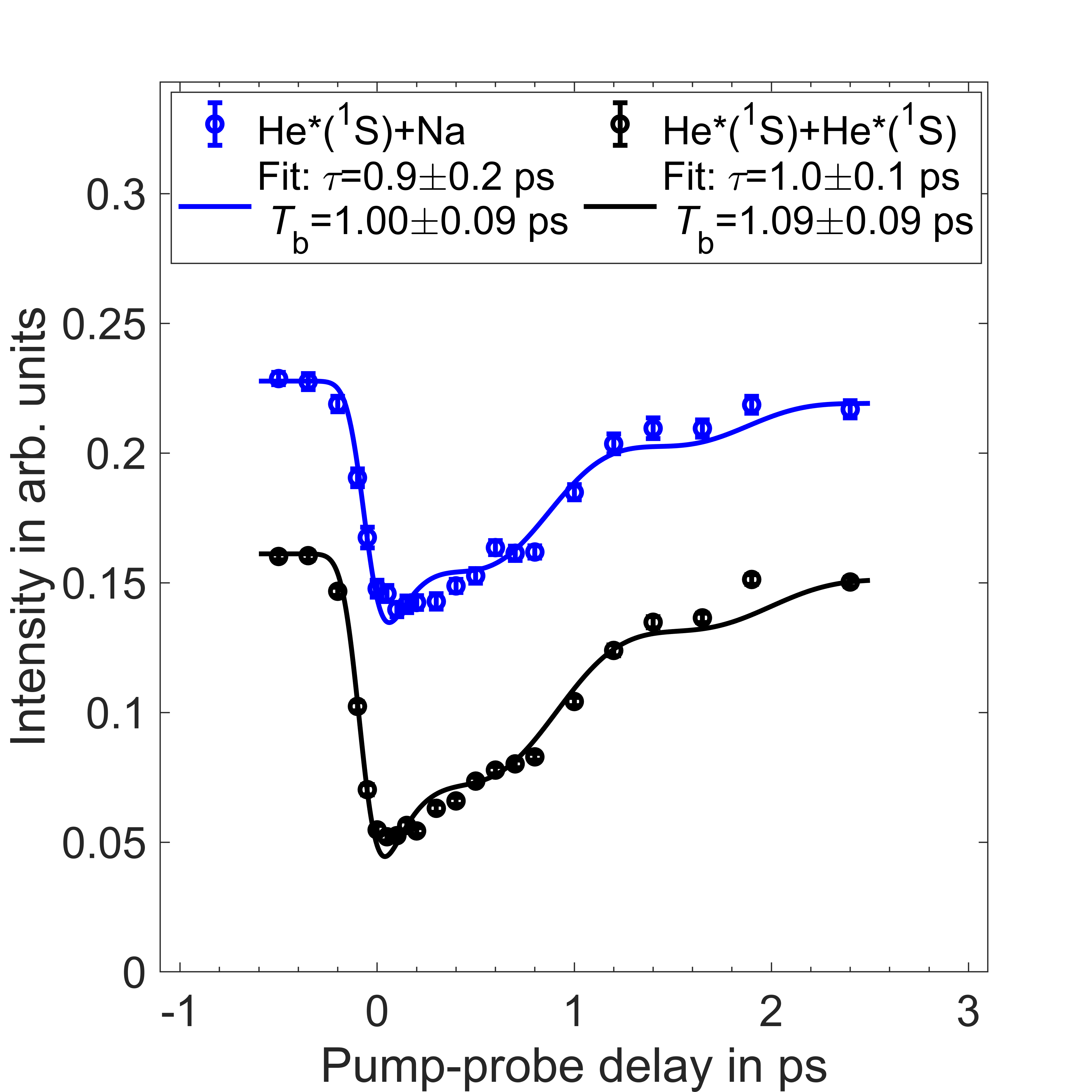}
	\caption{Electron yield of the two most dominant channels displayed in figure \ref{depletion} fitted with the extended fit model including an oscillatory term to account for the oscillation of the bubble around the excited He atom. The error bars on the data points show the standard deviation from the least-square regression of the multi-Gaussian fit model, and the resulted ICD lifetimes and bubble oscillation period from the extended exponential decay model are given in the legend with standard deviation. }
	\label{depletion_advanced}
\end{figure}
\begin{figure}[t]
	\centering
	\includegraphics[width = 0.95\columnwidth]{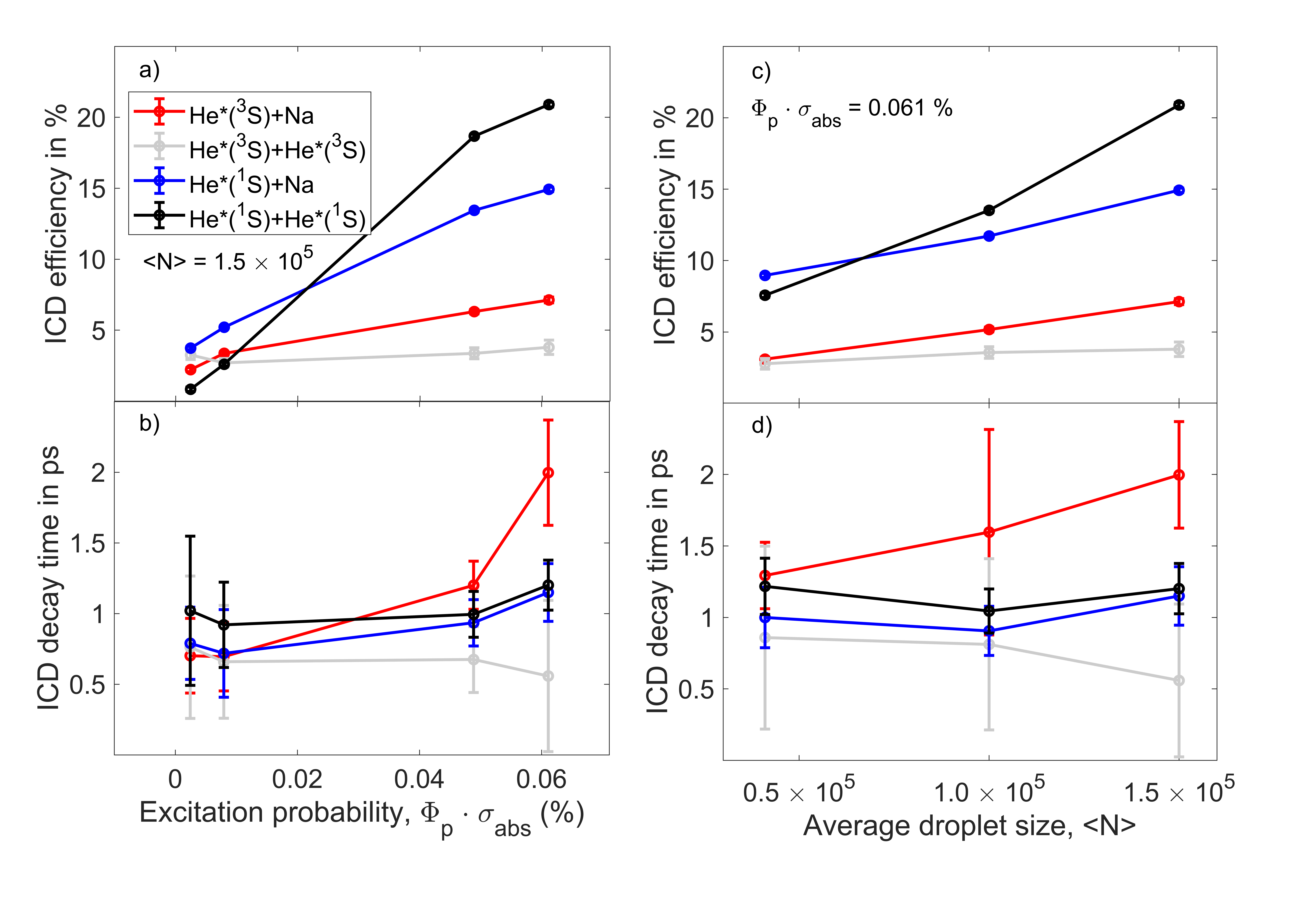}
	\caption{The ICD efficiency and lifetime determined from the variation of the excitation probability (a+b) and the average droplet size (c+d). The excitation probability is given as the probability for a single He atom in the droplet to be excited and is varied by changing the FEL intensity in the range $4\times10^8$~-~$1\times10^{10}$~Wcm$^{-2}$. The average droplet size was varied by cooling the He valve temperature from 18 to 14~K. All errorbars are showing standard deviation from least-square regression.}
	\label{ICD}
\end{figure}
\begin{figure}[h]
	\centering
	\includegraphics[width = 0.6\columnwidth]{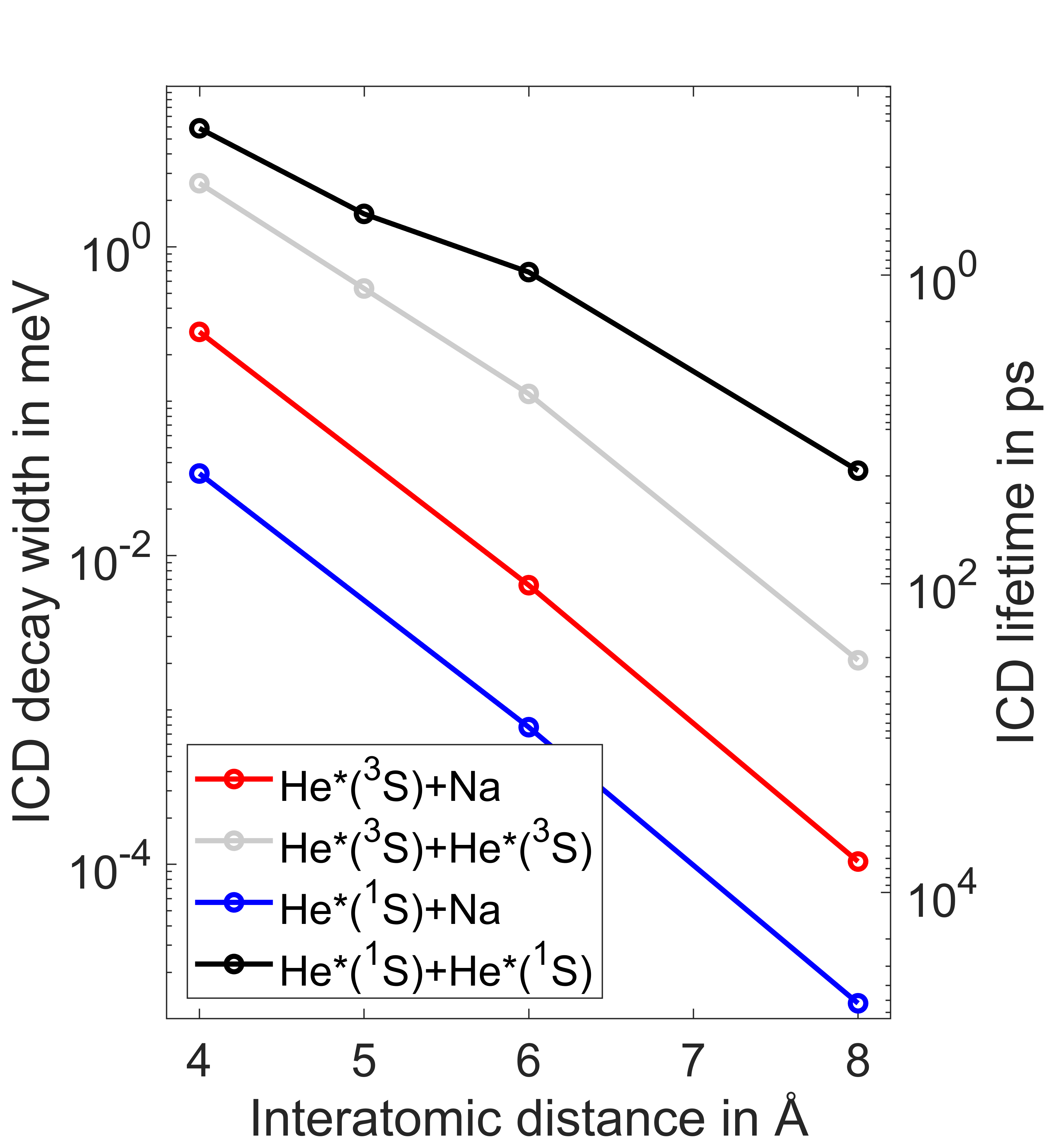}
	\caption{Calculated ICD decay width for the identified 4 main ICD channels in the experiment at different interatomic distances. Calculations are done using the Fano-CI-Stieltjes method~\cite{Miteva2017}. The He*+He* decay widths have previously been published in ref. \citenum{Laforge2021}.}
	\label{decay_width}
\end{figure}

To determine the time constants of the four main ICD channels, we varied the delay between the XUV pump pulse exciting the He droplet and a UV-probe pulse which ionizes the excited states and thereby depletes the ICD signal. Figure \ref{TRPES} shows a time-resolved PES (TRPES) for an average droplet size of $1.5\times10^5$ atoms per droplet, a FEL intensity of $1.0\times 10^{10}$~Wcm$^{-2}$ and a UV probe intensity of $2.6\times10^{11}$~Wcm$^{-2}$. At higher intensities of the UV probe, above threshold ionization (ATI) of higher order becomes dominant in the spectra~\cite{Michiels2021b}. The TRPES show that the ICD signal is depleted as excited atoms are increasingly photoionized from the initially excited $1s3p$/$1s4p\,^1P$ droplet state  by the UV-probe laser (1+1' ionization) around zero pump-probe delay (bright spot at 2.4~eV electron energy). At delays between 0 and 1~ps (XUV first, UV second), excited He droplets relax into the metastable $1s2s~^{1,\,3}$S atomic states giving rise to the two peaks at 1.6~eV and 2.4~eV by 1+2' photoionization. For further analysis of the low-energy part of the TRPES, we refer to our recent paper on the relaxation dynamics of superexcited He droplets~\cite{Asmussen2021}. In this phase, ICD begins to outcompete photoionization and ICD electrons reappear. The time constant of ICD can be determined from the reappearance time following depletion~\cite{Laforge2021}. For every pump-probe delay step, we determine the peak area of each of the four main ICD channels and fit the following depletion model for each channel
\begin{equation}
I(t) = I_0 - \frac{1}{2}~A~ \text{erfc} \left( \frac{\sigma^2-\tau(t-t_0)}{\sqrt{2} \sigma \tau} \right) 
\times~\text{exp} \left( \frac{-(t-t_0)}{\tau}\right).
\end{equation}      
Here, $\sigma$ is determined from the cross-correlation of the two laser pulses and $\tau$ is the ICD time constant. This function results from the convolution of the true signal function with the cross-correlation function and is consistent with a more extensive model for the pump-probe dynamics of ICD processes~\cite{Fasshauer2020,Laforge2021}. Figure \ref{depletion} shows the depletion curve and fit to the same data as displayed in figure \ref{TRPES}. The depletion model reproduces the experimental measurements to a large extent. However, the experimental measurements feature a kink in the rising edge around $1$~ps which is present in all pump-probe scans (see the supplementary material). The appearance time of $\sim$1~ps matches with the oscillation period of a bubble forming around an excited He atom in the droplet as determined from TDDFT calculations~\cite{Mudrich2020}. A similar oscillation of the bubble was reported for excited silver\cite{Mateo2013} and indium\cite{Thaler2018} atoms embedded in He nanodroplets with oscillation periods of $\sim$20~ps and $\sim$30~ps, respectively. In the case of indium-doped He nanodroplets, it could be seen that the photoelectron energy arising from ionization of the excited indium atom in the droplet had a small shift ($\sim$10~meV) to higher energy after an oscillation period. Similar behavior cannot be observed in our PES due to the proximity of the four ICD peaks in the spectra. However, as the bubble oscillates, the interatomic distance between the two excited He atoms or the Na cluster and the one excited He atom likely oscillates as well, thereby causing an oscillation of the probability for the pair of atoms to undergo ICD. 
To test this conjecture, we augment the fit model by a sine function that is superimposed on the exponential rise, where the period of the oscillation is treated as a free fit parameter. Assuming that the bubble oscillation has a lifetime longer or of the same order as the ICD lifetime, the oscillatory signal appears damped on the timescale of the ICD dynamics in the pump-probe measurement. 
Thus, the extended fit model is
\begin{equation}
I(t) = I_0 -\text{erfc} \left( \frac{\sigma^2-\tau(t-t_0)}{\sqrt{2} \sigma \tau} \right)  
\times~ \bigg[ \frac{1}{2}~A~\text{exp} \left( \frac{-(t-t_0)}{\tau}\right) 
\left(1 - B~ \sin \left(\frac{2\pi(t-t_0)}{T_b} \right)\right) \bigg],
\end{equation}        
where $T_b$ denotes the period of the bubble oscillation. Figure \ref{depletion_advanced} shows the fit to the two most dominant ICD channels (from He* in the $^1$S state) shown in figure \ref{depletion} with the extended fit model. The experimental data including the kink at $\sim$1~ps are reproduced very well, whereas the exponential lifetime remains essentially unchanged when using the extended fit model. Nevertheless, for the sake of keeping the fit model simple and for a direct comparison with our previous work~\cite{Laforge2021}, we continue to apply the purely exponential fit model in the following. To unambiguously confirm the validity of the oscillatory fit model, detailed three-dimensional simulations of the system of a Na-doped excited He nanodroplet would be needed, which falls beyond the scope of this work, though.

Figure \ref{ICD} shows the ICD efficiency and time constant for the four different channels for varying droplet size and excitation probability of a He atom in the droplet. The droplet size was controlled by varying the He expansion temperature in the range 14-18~K. The excitation probability was calculated as the product of the photon flux ($\Phi_{\text{p}}$) and the cross section ($\sigma_{\text{abs}}$) for exciting an atom in the droplet to the band associated with the atomic $1s3p$ and $1s4p$ configurations (2.9~Mb per He* atom)~\cite{Buchta2013,Joppien1993}. By varying the FEL intensity in the range $4\times10^8$~-~$1\times10^{10}$~Wcm$^{-2}$, the excitation probability is varied between 0.02\,\% and 0.6\,\%. The ICD efficiency is determined as the yield of ICD electrons at negative pump-probe delay (UV first) divided by the average number of excited states per droplet (in the case of He*+He* ICD, a factor of 2 is included to take into account the fact that two excited states are required). The latter is determined from the yield of photoelectrons from direct photoionization out of the droplet $1s3p$/$1s4p$ band at 0~ps pump-probe delay assuming the cross section for photoionization of the excited state in the droplet is equal to the atomic photoionization cross section for the $1s3p\,^1P$ state (3~Mb)~\cite{Chang1995}. 

We start our discussion of the results by looking at the ICD lifetime for different excitation probabilities and droplet sizes (figure \ref{ICD}b+d). The effective decay times are similar ($\sim$1~ps) for all four ICD channels at the different experimental conditions and match the decay time that we found for ICD in pure He droplets multiply-excited to the $1s2s\,^1S$ state at $h\nu = 21.6$~eV~\cite{Laforge2021}. There, it was found that the decay time was only weakly dependent on the excitation probability up to a much higher excitation probability ($\Phi_{\text{p}} \cdot \sigma_{\text{abs}} = 1$\%) than what we probe here. We compare the effective ICD time constant determined from our fit model with \textit{ab initio} calculations of the ICD width in He*-He* and He*-Na dimers obtained by the Fano-CI-Stieltjes method~\cite{Miteva2017}. The calculations predict that the ICD time constants vary over orders of magnitude as a function of interatomic distance. In He droplets this translates into a pronounced dependence on the mean density of excitations for the four channels, which is at odds with the experimental finding.

An ICD time constant of $\sim$1~ps corresponds to an interatomic distance $\lesssim 6$~Å, which is inconsistent with the density of excited states in the droplet in the experiment. For a droplet of size $4.1\times10^4$~atoms irradiated with a FEL pulse intensity of $10^{10}$~Wcm$^{-2}$, the number of excited states in the droplet is $\sim$25 corresponding to an average interatomic distance of $\sim$4~nm~between two excited states. This discrepancy supports our interpretation that ICD in He droplets is predominantly driven by the ultrafast dynamics of the superfluid medium surrounding the excited states instead of the long-range forces acting between the He* and the Na atoms, similarly to the case of pure He droplets~\cite{Laforge2021}. Our results further show that the measured ICD dynamics is largely independent of the electronic state of the droplet that is excited, as ultrafast relaxation brings the droplets to the lowest metastable states $1s2s\,^{1,\,3}$S prior to ICD. From our investigation of the relaxation dynamics, it was found that the complete relaxation to the two lowest metastable states happens on a similar time-scale as ICD, but that the electronic (interband) relaxation is faster, and thus happens before the He superfluid dynamics sets in~\cite{Asmussen2021}. ICD of dopants appears to be driven by a similar medium-dominated dynamics as the He*+He* ICD process in pure droplets.

Figure \ref{ICD}a+c shows the ICD efficiency for varying excitation probability and droplet size. For He*+He* ICD, the condition determining whether ICD takes place or not is that two excited states are formed in such close proximity that the two bubbles formed around them can merge and thereby push the two excited states close to one another~\cite{Laforge2021}. Similarly, for He*+Na ICD to be efficient, it is required that an excited state is formed within the distance of one bubble radius from the Na cluster located at the droplet surface. Based on this simple picture, one would expect the He*+He* ICD efficiency per He* excitation to increase as the He* density increases because it requires two He* excited states formed close to one another to induce ICD~\cite{Laforge2021}. In contrast, the He*+Na ICD efficiency per He* should remain constant with increasing He* density as the probability of exciting a He* near Na as opposed to exciting at He* elsewhere in the droplet is constant. 

In figure \ref{ICD}a, we see that the efficiency of He*+He* ICD indeed increases with rising He* density. However, the He*+Na ICD efficiency also rises, though at a lower rate than for the He*+He* counterpart. In fact, all ICD channels increase in efficiency both as a function of excitation probability and as a function of droplet size, see Figure \ref{ICD}c. According to our simple model, the He*+He* ICD efficiency is expected to be independent of the droplet size as the density of excited states is unchanged. In contrast, the He*+Na ICD efficiency is expected to decrease with droplet size as the surface-to-bulk volume ratio decreases and He*+Na ICD is a surface-selective process. 
This dependence of the ICD efficiency on the droplet size is likely due to the additional dynamics of He* excited atoms being ejected from the droplets when formed near the droplet surface~\cite{Mudrich2020,Laforge2021}. 
The same mechanism likely applies to the He*+Na system to some extent. However, as the Na dopant is located at the droplet surface, the migration of He* toward the droplet surface might also have the opposite effect of enhancing He*+Na ICD, which might explain the weaker droplet-size dependence of He*+Na ICD. More detailed three-dimensional model calculations as in~\citep{Mudrich2020} would be needed to assess the role of the He* ejection dynamics in this system.

It is interesting to compare the efficiency of He*+He* ICD found here for the $3p/4p$ excitation of the droplets with previous measurements for excitation into the lower $1s2p$ band at a similar He* density~\cite{Laforge2021}. We find that the efficiency is twice higher while the pump-probe dynamics is nearly identical. The increased efficiency may be due to stronger repulsion of the droplet environment from the more extended $3p/4p$ orbital prior to electronic relaxation~\cite{vonHaeften2011}. Accordingly, the bubbles forming around the $3p/4p$ He* excitations are transiently more extended, thereby enhancing ICD.

However, even when assuming that a twice larger bubble forms around the initially excited $3p/4p$ state prior to electronic relaxation~\cite{vonHaeften2011}, the effective He*-Na pair distance would not be sufficiently long to match the measured He*-Na ICD efficiency assuming only those He* located at close distance from the Na dopant contribute to the He*+Na ICD. For a mean droplet radius of $\sim12$~nm ($\left< N \right> \approx 1.5\times10^5$), a He* atom with a distance up to $\sim8$~nm would have to be allowed to contribute to ICD to achieve the measured $\sim10$\,\% efficiency (see supplementary material). This distance is more than an order of magnitude larger than the bubble radius (6.8~\AA~for the $2p\,^1S$ state~\cite{Asmussen2021}), which we would expect to be the maximum distance for ICD to happen. Thus, the experimental results show that ICD is both fast and efficient, which is inconsistent with our simple geometric model since fast decay requires short interatomic distances which would imply a small active volume of the droplet leading to ICD. 

Our simple model is based on the excitation being localized on a single atom, which \textit{ab initio} calculations have shown to occur within $\sim$100~fs for small He clusters ($N=7$)~\cite{Closser2014a}. This behavior is different compared to heavier rare gas clusters where excitations remain delocalized in excitonic states~\cite{Buchenau1991}. In our model, we expect the excitation to localize on a random He atom, leading to a homogeneous density of excited states in the droplet. However, one may expect that just after the excitation, when the state is still delocalized and exciton-like, the excitation migrates through the droplet towards the dopant due to the dopant's high polarizability compared to He giving rise to attractive dispersion forces. Similarly, in the case of two excitations in the droplet, one would expect that interaction between the two excitons would favor their localization near one another. Detailed quantum calculations are needed to confirm the hypothesized behavior in the early excitonic state. Assuming this model applies, it would readily explain the increased efficiency of ICD when exciting into a higher-lying droplet band as the initial pre-localized state would be more delocalized and thereby more likely to overlap with a dopant or another exciton. 
After the ultrafast localization has taken place, the ICD dynamics is driven by the dynamics in the superfluid medium (bubble formation), which explains why the effective lifetime of ICD is independent of excitation energy. 

Another reason for an inhomogeneous distribution of excited states may be the selective deposition of energy by the XUV-laser at the surface of the droplets. Calculations of the perturbed Rydberg state energy in the He droplet environment showed that at our pump-pulse photon energy the spectral overlap with the $1s3p\,^1P$ atomic state is largest near the surface of the droplet~\cite{Kornilov2011a}. This could cause an enhanced excitation density at the droplet surface where the Na dopant cluster is located.

In summary, we have measured for the first time the dynamics of ICD between an excited He atom and a Na dopant atom in superexcited He nanodroplets. Four main channels were identified involving the lowest metastable atomic states $1s2s\,^{1,\,3}$S following relaxation from the initial highly-excited droplet state. Similar ICD lifetimes $\sim 1$~ps are determined for the four ICD channels indicating that the dynamics is mainly driven by the dynamical response of the droplet environment. Furthermore, the efficiencies of the four ICD channels are high compared to previous measurements on pure He nanodroplets excited at lower photon energy, and a simple model assuming a homogeneous distribution of localized excited states could not account for the high efficiency of He*+Na ICD. These results pave the way for further time-resolved measurements and simulations of ICD-mediated single or double ionization processes of molecules and clusters~\cite{Laforge2016c,Laforge2019a,BenLtaief2020}. 

\section{Experimental methods}
The experiments were performed at the low density matter (LDM) endstation of the seeded FEL FERMI in Trieste, Italy~\cite{Lyamayev2013a} equipped with a magnetic bottle electron spectrometer~\cite{Squibb2018}. The FEL was tuned to the center position of the $1s3p/1s4p$ band ($h\nu = 23.7$~eV) by seeding it with the third harmonic of a Ti:Sapphire laser (261~nm) and setting the undulators to the 5$^{th}$ harmonic of the seed laser~\cite{Allaria2012c}. The temporal duration of the FEL pulses was 80~fs (FWHM), and their intensity was varied using different filters in the range $4.0\times10^8$~-~$1.7\times10^{10}$~Wcm$^{-2}$. The intensity profile of the beam in the interaction region was nearly Gaussian with a FWHM of $70~\mu$m. The UV probe laser pulses were generated from the second harmonic of the Ti:Sapphire seed laser ($h\nu' = 3.2$~eV), and the pulse intensity was $2.6\times10^{11}$~Wcm$^{-2}$ for all measurements. The laser intensities were determined from the pulse energies and intensity distributions measured by dedicated monitors. They were corrected for the calculated transmission of the beamline~\cite{Svetina2015a}. The cross correlation of the pump and probe pulse was experimentally determined to be 170~fs FWHM. 

He nanodroplets were formed by expanding He gas from a high-pressure reservoir (50~bar) through a pulsed, cryogenically cooled Even-Lavie-type valve at a pulse repetition rate of 50~Hz~\cite{Pentlehner2009a}. The mean size of the droplets was varied in the range of $4 \times 10^4$~-~$7 \times 10^5$ atoms per droplet by varying the valve temperature between 10 and 18~K. After formation, the He droplets passed through a 1~cm long heated cell filled with Na where each droplet picked up a variable number of Na atoms according to the Na vapor pressure.

\begin{acknowledgement}
	The authors are grateful for financial support from the
	Deutsche Forschungsgemeinschaft (DFG) within the project
	MU 2347/12-1 and STI 125/22-2 in the frame of the Priority
	Programme 1840 QUTIF, and from the Carlsberg Foundation.
	R. F. thanks the Swedish Research Council (VR) and the Knut
	and Alice Wallenberg Foundation for financial support.	
\end{acknowledgement}

\bibliography{He_Na_ICD}
\end{document}